\documentclass[%
 reprint,
nofootinbib,
 amsmath,amssymb,
 aps,
]{revtex4-1}

\usepackage{graphicx}% Include figure files
\usepackage{dcolumn}% Align table columns on decimal point
\usepackage{bm}% bold math
\usepackage{hyperref}
\hypersetup{colorlinks=true, citecolor=blue, linkcolor=blue, urlcolor=blue}
 \usepackage{multirow}
 \usepackage[table]{xcolor}

\def\sfrac#1#2{{\textstyle{#1\over #2}}}

\begin{document}

\preprint{APS/123-QED}

\title{Simple quintessence models in light of DESI-BAO observations}

\author{James M. Cline}
\email{jcline@physics.mcgill.ca}
\thanks{ORCID: \href{https://orcid.org/0000-0001-7437-4193}{0000-0001-7437-4193}}
\affiliation{McGill University Department of Physics \& Trottier Space Institute, 3600 Rue University, Montr\'eal, QC, H3A 2T8, Canada}

\author{Varun Muralidharan}
\email{varun.muralidharan@mail.mcgill.ca}\thanks{ORCID: \href{https://orcid.org/0009-0004-8460-5230}{0009-0004-8460-5230}}
\affiliation{McGill University Department of Physics \& Trottier Space Institute, 3600 Rue University, Montr\'eal, QC, H3A 2T8, Canada}

\begin{abstract}
Recent analyses from the DESI collaboration suggest that the dark energy density of the Universe may be decreasing with time, slowing the acceleration of the scale factor $a$.   Typically these studies are performed assuming an ansatz for
the equation of state $w(a)$.  In this work, we instead consider simple models of a scalar quintessence potential with linear and quadratic behavior, which could be more representative of real models than particular parametrizations of $w(a)$. We observe a significant preference for dynamical dark energy when using supernova data from DESY5 along with DESI BAO and Planck data, at the cost of slightly exacerbating the Hubble tension. However, when using supernova data from Pantheon+ or Union3, we find only a mild preference for dynamical dark energy.

\end{abstract}

\maketitle

%\tableofcontents

\section{Introduction}

The standard $\Lambda$CDM model of cosmology is under intense scrutiny as data continue to improve.  There are intriguing hints that the theory may be challenged, but so far no fully convincing evidence.  The Dark Energy Spectroscopic Instrument (DESI) \cite{DESI:2025fii} has rekindled interest in models of evolving dark energy, as opposed to the cosmological constant \cite{Ramadan:2024kmn, Akrami:2025zlb, Berbig:2024aee, deSouza:2025rhv, Ling:2025lmw, Chaussidon:2025npr, Kessler:2025kju, Moffat:2025jmx}.  

It is common to adopt a seemingly model-independent approach to characterize the time evolution of the dark energy, by parametrizing its equation of state $w = P/\rho$ as some simple function of the redshift $z$, and constraining the extra parameters associated with $w(z)$ \cite{Wolf:2025jlc, Shlivko:2025fgv}.  However, it is not obvious that such parametrizations are good representations of specific models, which might be theoretically well-motivated.

One theoretical motivation is the idea that the true ground state of the Universe has negative cosmological constant, but is temporarily experiencing positive dark energy due to the displacement of the quintessence field from its minimum.  Ref.\ \cite{VanRaamsdonk:2023ion} studied such models in the context of the AdS/CFT correspondence (or
more generally string theory), which manifests a preference for negative cosmological constant.  Construction of consistent string theory vacua is relatively straightforward in an anti-deSitter background, while attempts to lift $\Lambda$ to positive values \cite{Kachru:2003aw,Balasubramanian:2005zx} remain uncertain, in terms of all approximations being under control, in the string theory context \cite{Chakravarty:2024pec,Schachner:2025vol,McAllister:2023vgy}.
 A negative cosmological constant with an evolving component of dark energy has been investigated by several authors recently \cite{Sen:2021wld, Mukherjee:2025myk, Wang:2025dtk, Muralidharan:2024hsc}. 

In this study, we consider simple, generic quintessence potentials with respect to the latest cosmological data.  These models are not meant to be ultraviolet complete, but rather to provide a good approximation to a scalar field (quintessence) potential $V(\phi)$ during the limited epoch of redshifts $z\in [0,2.3]$, for which current data give hints of evolving dark energy.  It amounts to Taylor-expanding $V(\phi)$ around its initial value using a small number of terms.  This is a familiar procedure in the literature, but it is worth updating in light of the recent data.  Moreover, we wish to investigate which ``model-independent'' parametrizations of $w(z)$ are well-suited to these kinds of quintessence models.

We remind the reader about the basic framework for quintessence in Section \ref{SPsect}, while introducing the restricted classes of potentials we will study, referred to as linear, hilltop, and hill-bottom.  In Section \ref{sec:anal} we present the numerical analysis of the models with respect to
current cosmological data.  Discussion and conclusions are given in Section \ref{sec:disc}.

\section{Scalar Potential}
\label{SPsect}
Quintessence scalar fields $\phi$, minimally coupled to gravity, provide an alternative to the cosmological constant as models of dark energy.  The evolution of the field due to the scalar potential $V(\phi)$ is described by the classical field equation,
\begin{equation}
    \ddot{\phi} + 3H\dot{\phi} + V'(\phi) = 0\,, 
\end{equation}
where $H=\dot{a}/a$ is the Hubble parameter, and the dot denotes a cosmic time $t$ derivative. The energy density of the field $\rho_{\phi}$ is defined by $\rho_\phi = \frac{1}{2}\dot{\phi}^2 + V(\phi)$, which gives the modified Friedmann equation
\begin{equation}
    H^2 = \frac{8 \pi G}{3}\left(\rho_m + \sfrac{1}{2}\dot{\phi}^2 + V(\phi)\right)\,.
\end{equation}
Here $\rho_m$ is the matter density;  we ignore contributions of radiation since we are interested in low redshifts, and we assume the spatial curvature of the Universe is negligible.

This work focuses on ``thawing'' quintessence models \cite{Tsujikawa:2013fta}, in which the field $\phi$ starts from rest in the early universe. This is a natural assumption if it is initially frozen due to the Hubble damping term, resulting in a cosmological constant-like equation of state $w = p_\phi/\rho_\phi \simeq -1$. With time, the Hubble parameter decreases and the field starts rolling down the potential.

We  model the shape of the potential when the field starts rolling using a linear or a quadratic approximation, 
including potentials that can reach negative values in the future,  like the ones Ref.\ \cite{VanRaamsdonk:2023ion} consider. This is motivated the observation that effective field theories with well-understood UV competitions
(such as from supergravity or string theory) typically have cosmological constant $\Lambda \leq 0$.  Taylor expanding the potential about its initial value gives
\begin{equation}
    V(\phi) = V_0 + V_1\phi + \sfrac{1}{2}V_2\phi^2 +\mathcal{O}(\phi^3),
\end{equation}
where $V_0=\Lambda$, $V_1$ and $V_2$ are constants. The simplest case is the linear potential, in which $V_2=0$. 
The sign of $V_1$ is irrelevant, since it only determines in which direction the field rolls. 

In the quadratic case, we can eliminate the linear term by doing a field redefinition $\phi\to\phi\ +$ constant.
The behavior of the system depends qualitatively on the 
the sign of $V_2$. We dub the $V_2>0$ case as a ``hill-bottom'' potential, 
which has a global minimum.  The field will eventually come to rest there
at some time in the future.  Depending on the sign of $V_0$, this will be accompanied by eternal acceleration or alternatively a big crunch.
The $V_2 <0$ case is known as a hilltop or tachyonic potential. The eventual
runaway of the field will inevitably lead to a big crunch in the future if this potential is taken literally; more realistically, the full potential could reach a minimum in the future, at field values beyond the region of validity of our Taylor expansion.  We are only concerned with the past and present observations in this work.

\section{Analysis}
\label{sec:anal}
To study the evolution of the field, it is convenient to replace the independent variable $t$ with the scale factor $a$. The scalar field equation of motion can be rewritten as coupled first-order equations,
\begin{equation}
    \begin{split}
        \frac{d\phi}{da} &= \frac{\pi_\phi}{aH}\,,\\
        \frac{d\pi_\phi}{da} &= -\frac{V'(\phi)}{aH} -  3\frac{\pi_\phi}{a}, \\
        H &= \sqrt{{8\pi G\over 3}\left(\rho_c\frac{\Omega_m}{a^3}+\frac{1}{2}\pi_\phi^2  +V(\phi)\right)}
    \end{split}
\end{equation}
where $\pi_\phi = \dot{\phi}$ is the canonical momentum of the field and
$\rho_c$ is the critical density of the Universe.
Going to geometrized units where $8\pi G/3=1$ and $c=1$, $\phi$ becomes dimensionless, while $\pi_\phi\sim \sqrt{\rho_c}$ and the potential parameters are of order $\rho_c$.  One can adopt $\rho_c$ as the unit of energy density and thereby set it to unity (in addition to $8\pi G/3$) to make the equations of motion and $H$ dimensionless.
It is straightforward to transform the various quantities back to physical units, if desired, 
using appropriate powers of $\rho_c$ and $8\pi G/3$.

Ref.\ \cite{VanRaamsdonk:2023ion} evolves the equations backwards in time starting from today ($a=1$), assuming some value of $\phi(1)$,  
and determining $\pi_\phi (1)$ from $H_0$ and $\Omega_m$. However, evolving the equation backwards in time often leads to solutions that diverge at large redshifts, which presents an obstacle to Monte Carlo sampling of the parameter space. 
Instead, we evolve the equations forward starting from rest ($\pi_\phi=0$) at 
an early epoch $a_i=0.001$. One can easily determine the initial value $\phi (a_i)$ that leads to the observed present value of  $\Omega_{\rm DE}$,  the fraction of energy density in dark energy.

To assess the models' compatibility with current data, 
we incorporated the evolution history of the scalar field energy density $\rho_{\phi}$ into the Boltzmann code \texttt{CAMB} \cite{Lewis:1999bs}, using the parametrized post-Friedmann approach \cite{PhysRevD.78.087303} to calculate dark energy perturbations. To scan the parameter space, we used the publicly available Markov Chain Monte Carlo sampler \texttt{Cobaya} \cite{Torrado:2020dgo}. Convergence of the chains was monitored using the Gelman-Rubin statistic \cite{Gelman:1992zz} requiring $R-1 \leq 0.02$. We considered the following data in our analysis:

\begin{itemize}
    \item \textbf{BAO:} Baryon acoustic oscillation measurements of distances,
    using Data Release 2 of DESI \cite{DESI:2025zgx};
    \item  \textbf{SN:}  Type Ia supernova distance determinations, using
    the Dark Energy Survey year 5 (DESY5) data release \cite{DES:2024tys}, Union3 \cite{Rubin:2023ovl} and Pantheon+ \cite{Brout:2022vxf}.
    \item  \textbf{DES}: Determination of clustering and cosmic shear using
    year-1 data from the Dark Energy Survey (DES)
    \cite{DES:2017myr, DES:2017tss, DES:2017qwj}; 
     \item \textbf{Planck:}  Anisotropies of the cosmic microwave background
     (CMB) polarization and temperature from Planck 2018 \cite{Planck:2018vyg}, as well as cross-correlations;
     \item \textbf{Lensing:} The power spectrum of the lensing potential for the CMB, derived from the Planck 2018 \cite{Planck:2018lbu} four-point temperature correlator.
\end{itemize}

\begin{center}
\makeatletter

    \def\CT@@do@color{%
      \global\let\CT@do@color\relax
            \@tempdima\wd\z@
            \advance\@tempdima\@tempdimb
            \advance\@tempdima\@tempdimc
    \advance\@tempdimb\tabcolsep
    \advance\@tempdimc\tabcolsep
    \advance\@tempdima2\tabcolsep
            \kern-\@tempdimb
            \leaders\vrule
    %^^A                     \@height\p@\@depth\p@
                    \hskip\@tempdima\@plus  1fill
            \kern-\@tempdimc
            \hskip-\wd\z@ \@plus -1fill }
    \makeatother
    
\renewcommand{\arraystretch}{1.6}
\begin{table}[h]
\begin{tabular}{|c|c|c|c|c|}
\hline

\multicolumn{1}{|c|}{Dataset} & Potential  & $\Delta\chi^2$ & AIC &  $H_0$ \\ \hline \hline
 {\cellcolor{red!25}} &  Linear & $-2.9$ & $+1.1$ &  $67.88^{+0.55}_{-0.46}$ \\ 
   {\cellcolor{red!25}}  & Hilltop & $-2.9$ & $+1.1$ & $67.59^{+0.58}_{-0.51}$ \\ 
   \parbox[t]{2mm}{\multirow{-3}{*}{\cellcolor{red!25}{\rotatebox[origin=c]{90}{Pantheon+}}}} & Hill-Bot & $-2.7$ & $+1.3$ & $67.66^{+0.61}_{-0.50}$  \\ \hline\hline
    
{\cellcolor{green!25}} & Linear & $-8.1$ & $-4.1$ & $67.17^{+0.66}_{-0.57}$ \\ 
  {\cellcolor{green!25}}     & Hilltop & $-10.0$ & $-6.0$ & $66.85^{+0.58}_{-0.58}$ \\ 
 \parbox[t]{2mm}{\multirow{-3}{*} {\cellcolor{green!25}{\rotatebox[origin=c]{90}{DESY5}}}}   & Hill-Bot & $-8.8$ & $-4.8$ & $66.92^{+0.57}_{-0.57}$ \\ \hline\hline

{\cellcolor{blue!25}} & Linear & $-3.2$ & $+0.8$ & $67.55^{+0.89}_{-0.50}$ \\ 
  {\cellcolor{blue!25}}   & Hilltop & $-3.6$ & $+0.4$ & $66.91^{+0.89}_{-0.74}$ \\ 
 \parbox[t]{2mm}{\multirow{-3}{*} {\cellcolor{blue!25}{\rotatebox[origin=c]{90}{Union3}}}}   & Hill-Bot & $-2.7$ & $+1.3$ & $67.13^{+0.87}_{-0.73}$ \\ \hline
    
\end{tabular}
\caption{Best-fit $\Delta\chi^2 = \chi^2 - \chi^2_{\rm \Lambda CDM}$, $\Delta$AIC = AIC $- \text{AIC}_{\rm \Lambda CDM}$ and the mean value of $H_0/(\text{km s}^{-1}\text{Mpc}^{-1})$ at 65\% confidence level (CL) for the 3 different potentials using 3 different choices of SN datasets. The full dataset used here is Planck + Lensing + DESI BAO + DES, and the supernova dataset mentioned.} 
\label{table:table1}
\end{table}

%Pantheon = 4703.6
%DESY5 = 4946.93
%Union3 = 3325.95
\end{center}

\begin{figure}
    \centering
    \includegraphics[height=8cm]{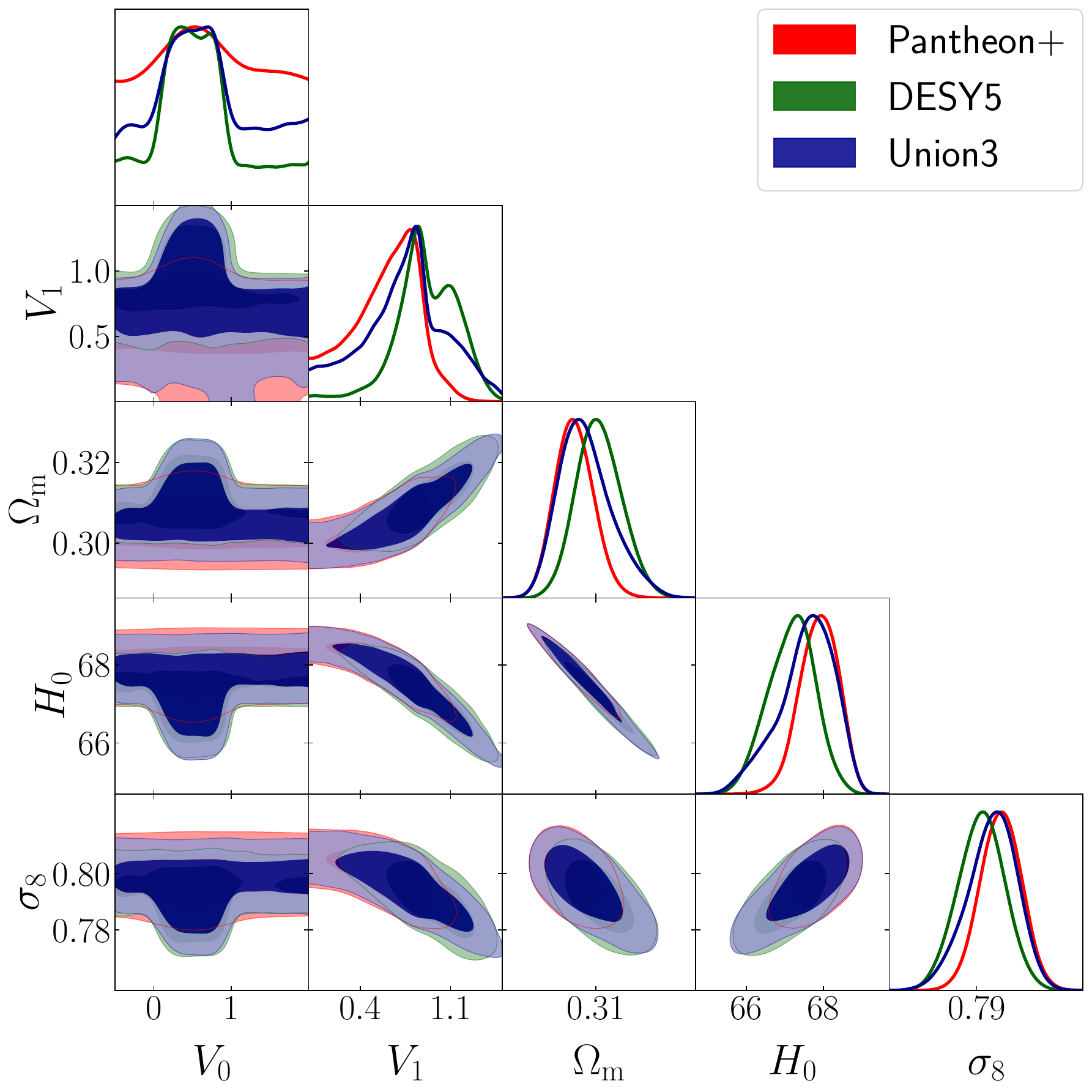}
    \caption{Marginalized posterior on $V_0, V_1, \Omega_m, H_0$ and $\sigma_8$ obtained in the full MCMC scan for the linear potential. The full dataset used here is Planck + Lensing + DESI BAO + DES, and the supernova dataset mentioned.}
    \label{fig:linear}
\end{figure}

Likelihoods were computed by combining the different datasets, except that we do not simultaneously use all the SN data, since some of them are in tension with each other so far as the determination of dark energy parameters is concerned.  In the following, the three possible combinations (nonSN data combined with one of the 3 SN sets) will be referred to using the name of the SN dataset. We adopt the same initial priors and proposal step sizes for $V_0, V_1$, and $V_2$ across chains.

\subsection{Linear Potential}
We start with the linear potential model, 
$V(\phi) = V_0 + V_1\phi$, assuming a uniform prior $\mathcal{U}[-0.5, 2.0]$ on $V_0$ and $\mathcal{U}[0, 1.5]$ on $V_1$.\footnote{Recall 
that $V_0$ and $V_1$ are in units of $\rho_c$ when $8\pi G/3=1$.  More generally, $V_n$ is in units of 
$\rho_c(8\pi G/3)^{n/2}$. }
The results of Monte Carlo sampling of the parameter space are shown shown in Table \ref{table:table1}: using the Pantheon+ supernova dataset shows that the model provides no significant improvement over $\Lambda$CDM.
 Fig. \ref{fig:linear} shows the 1D and 2D joint marginalized posterior on the free parameters $V_0$ and $V_1$ and the derived parameters $\Omega_m$, $H_0$ and $\sigma_8$. 

\begin{figure}
    \centering
    \includegraphics[height=6cm]{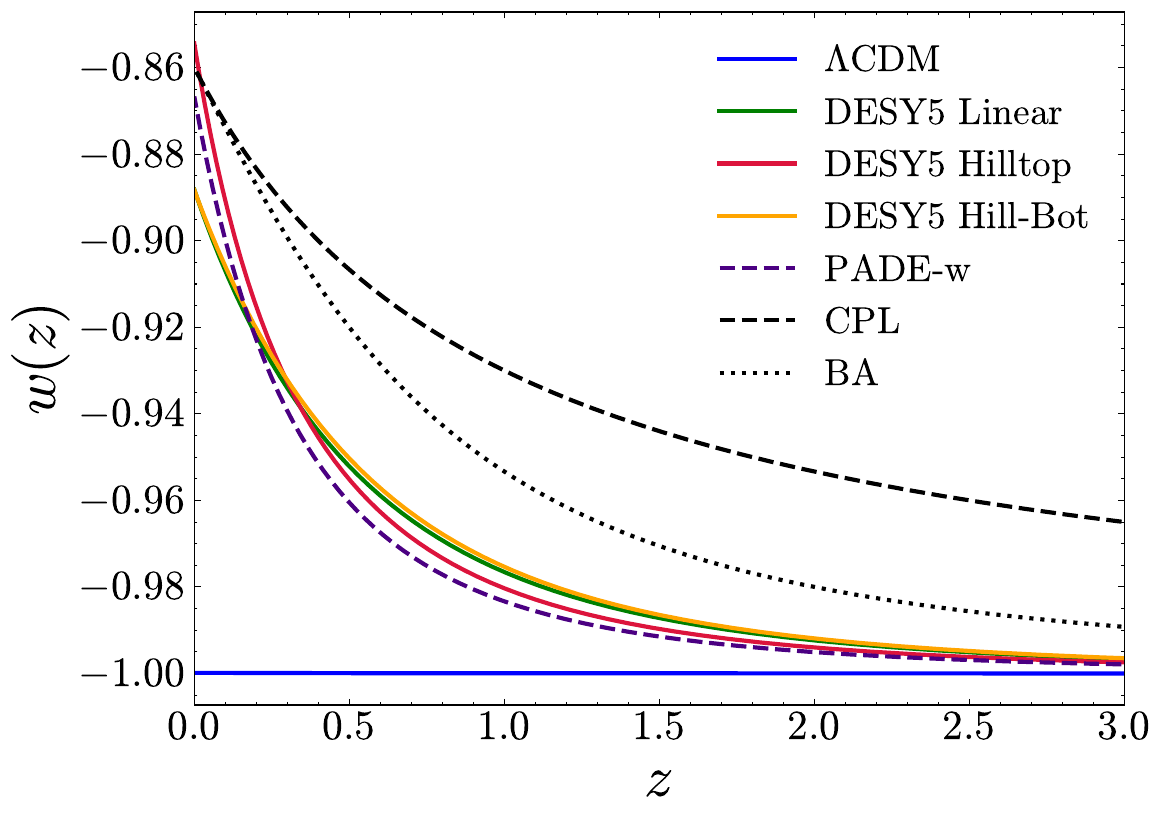}
    \caption{Equation of state of dark energy as a function of redshift for the Pantheon+ best fit ($\Lambda$CDM) and the DESY5 best-fit for the 3 different potentials compared to PADE-w parametrization with $\epsilon_0=0.2$ and $\eta_0=3$ and $w_0w_a$ parametrization with $w_0 = -0.86$ and $w_a = -0.14$.}
    \label{fig:wa}
\end{figure}

The Pantheon+ and Union3 datasets give only a marginal improvement  in $\chi^2$ with $\Delta\chi^2 = \chi^2 - \chi^2_{\rm \Lambda CDM}$ of $-2.9$ and $-3.2$ respectively. In contrast, the DESY5 SN dataset gives a strong preference for the linear potential over $\Lambda$CDM, with a $\Delta\chi^2$ of $-8.6$. To interpret the significance of these results, we employ the Akaike Information Criterion (AIC): a preferred model has lower $AIC = 2k - 2\ln(\mathcal{L}) = 2k + \chi^2$ where $k$ is the number of additional parameters in the theory. The linear models have two parameters,\footnote{Even though physically $V_0$ is equivalent to the cosmological constant, in practice it constitutes an additional parameter; in $\Lambda$CDM, $\Omega_\Lambda = 1 - \Omega_m$ is fixed.} $V_0$ and $V_1$. For the Pantheon+ and Union3 best fits, the AIC for $\Lambda$CDM is less than that of the models, hence $\Lambda$CDM is preferred. The modest decrease in $\chi^2$ for the model can be attributed  to adding new parameters. On the other hand, the DESY5 fit has AIC below that of  $\Lambda$CDM by $ 4.1$ which implies a strong preference. The best-fit value of $H_0$ for the DESY5 fit $67.17^{+0.66}_{-0.57} ~\text{km s}^{-1}\text{Mpc}^{-1}$ is almost consistent with the Planck measurement of $H_0 = 67.36 \pm 0.54$, which is at least $4.5\sigma$ away from the local distance ladder measurement of $H_0 = 73.2 \pm 1.3 ~\text{km s}^{-1}\text{Mpc}^{-1}$ \cite{Riess:2020fzl}. 

The best-fit values of $V_0=0.79$ and $V_1=-1.05$ for the DESY5 fit imply significant evolution of the dark energy in the recent past. The dependence of $w(z)$ for this case is shown in Fig. \ref{fig:wa}. For this model, $w$ will continue to increase in the future, halting the acceleration and tending toward the collapse of the Universe due to its negative cosmological constant.

\subsection{Hilltop}
\begin{figure}
    \centering
    \includegraphics[height=8cm]{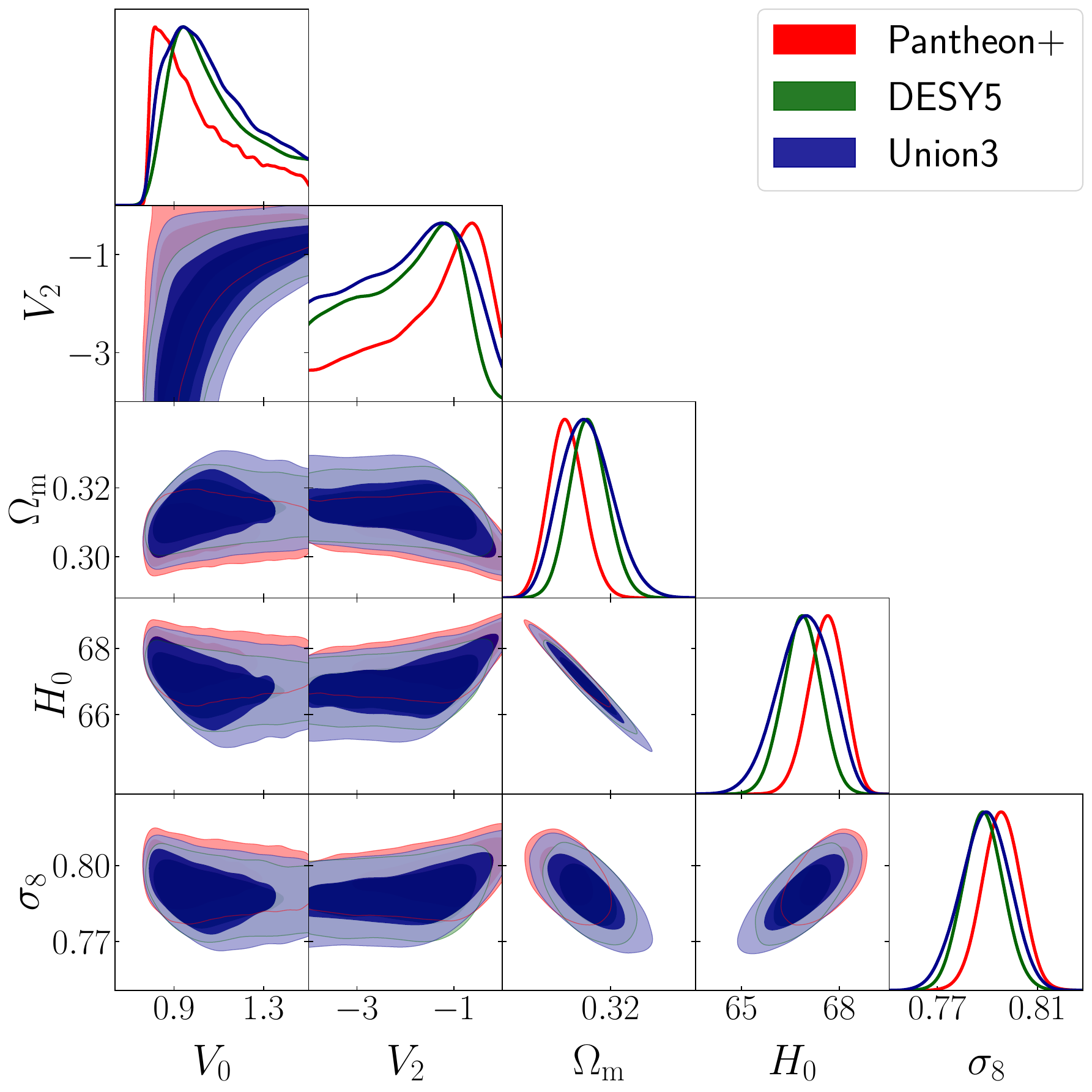}
   \caption{Marginalized posterior on $V_0, V_2, \Omega_m, H_0$ and $\sigma_8$ obtained in the full MCMC scan for the quadratic hilltop potential. The full dataset used here is Planck + Lensing + DESI BAO + DES, and the supernova dataset mentioned.}
    \label{fig:hilltop}
\end{figure}

Next, we consider quadratic hilltop potentials defined by $V(\phi) = V_0 + \frac{1}{2}V_2\phi^2$ where $V_2 < 0$. (Recall that in this case, a linear term can be absorbed removed through a shift in $\phi$). We take a uniform prior $\mathcal{U}[0.7, 1.5]$ on $V_0$ and $\mathcal{U}[-4,0]$ on $V_2$. Table \ref{table:table1} lists the $\Delta\chi^2$ and  $\Delta \mathrm{AIC = AIC - AIC_{\Lambda CDM}}$ for the hilltop potential for the 3 different choices of supernova datasets, and Fig. \ref{fig:hilltop} displays the marginalized posteriors on $V_0, V_2, \Omega_m, H_0$ and $\sigma_8$. 
Similar to the linear potential, the Pantheon+ and Union3 dataset yields an insignificant improvement with respect to $\Lambda$CDM with $\Delta\chi^2 = -2.9$ and $-3.6$ respectively. But again, the DESY5 dataset improves the fit substantially, with $\Delta \chi^2 = -10.0$. The best-fit values of $V_0=0.9$ and $V_2=-3.2$ lead to significant rolling of the quintessence field.

 Like the linear case, the quadratic model also has only two extra parameters; hence the AIC is higher for the Pantheon+ and Union3 fit than $\Lambda$CDM which implies that $\Lambda$CDM is still the preferred model. However, the AIC of the DESY5 fit is significantly lower than that of $\Lambda$CDM, indicating a strong preference for the quintessence model. We observe that the mean value of the Hubble parameter $H_0 = 66.85^{+0.58}_{-0.58}$, exacerbating the Hubble tension by about $0.4\sigma$.

\subsection{Hill-bottom}
\begin{figure}
    \centering
    \includegraphics[height=8cm]{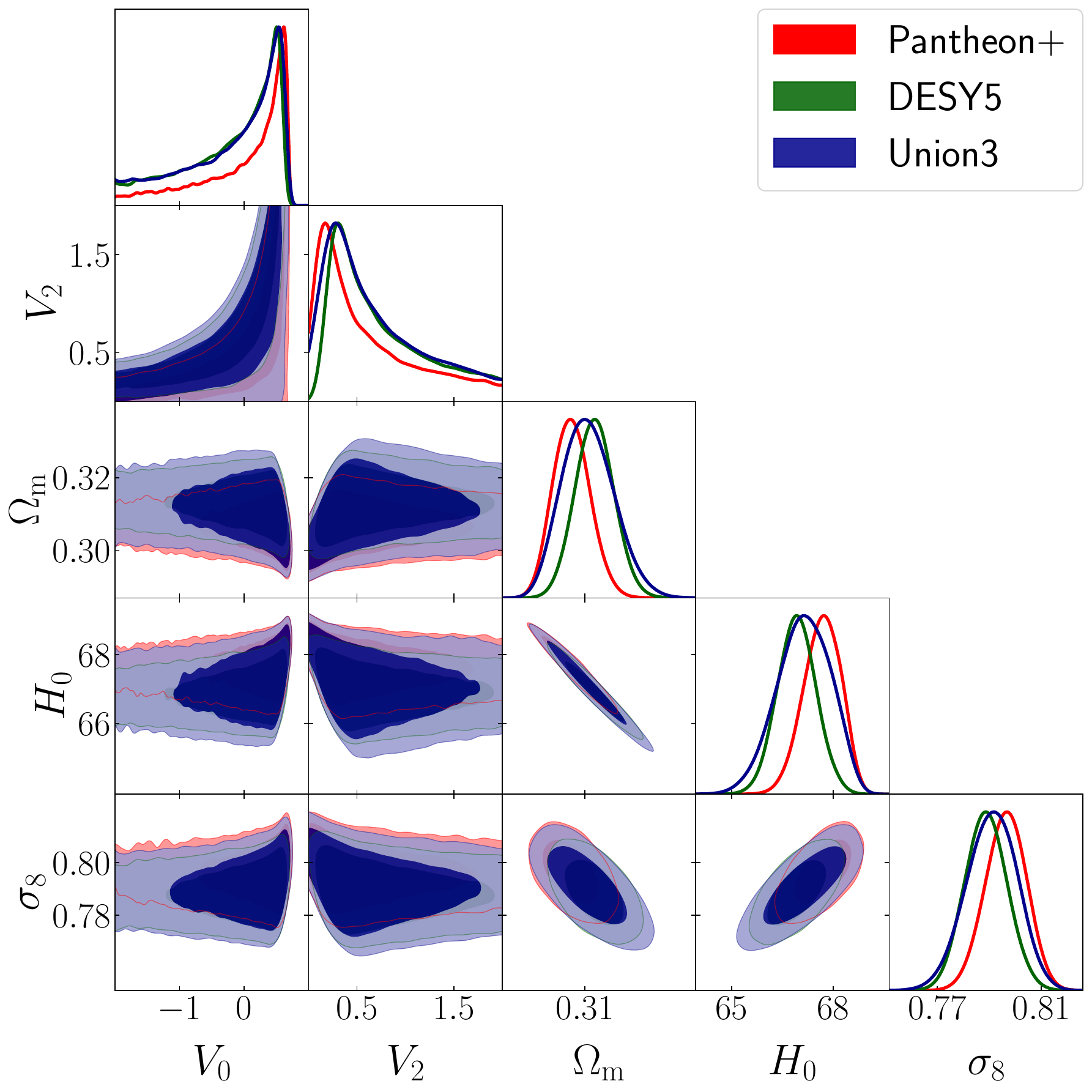}
   \caption{Marginalized posterior on $V_0, V_2, \Omega_m, H_0$ and $\sigma_8$ obtained in the full MCMC scan for the quadratic hill-bottom potential. The full dataset used here is Planck + Lensing + DESI BAO + DES, and the supernova dataset mentioned.}
    \label{fig:hillbot}
\end{figure}

Lastly, we study hill-bottom potentials, with $V_2 >0$, assuming a uniform prior $\mathcal{U}[-2,1]$ on $V_0$ and $\mathcal{U}[0,2]$ on $V_2$. The $\Delta\chi^2$ and $\Delta$AIC values for the model are also shown in Table \ref{table:table1} and the marginalized posterior on the parameters is shown in Fig. \ref{fig:hillbot}.  Like for the other models, the hill-bottom with Pantheon+ and Union3 only mildly improve the fit compared to $\Lambda$CDM with $\Delta \chi^2 = -2.7$ for both.  This is insignificant from the perspective of the AIC. With DESY5 there is a distinct local minimum in its $\chi^2$ relative to the hilltop, though not as deep as in that  model: the improvement in $\chi^2$
is 8.8.  The preferred Hubble parameter is pushed to more discrepant values relative to the CMB determination, by  $\sim 0.3\,\sigma$.

\subsection{Equation of state evolution}

We are interested to know how well various parametrizations of the dark energy equation of state $w(z)$ can match the behavior corresponding to the models.  
In Fig. \ref{fig:wa} we show the behavior of the equation of state of dark energy for the three different model classes, assuming the DESY5 dataset, which offers the greatest improvement relative to $\Lambda$CDM. The equation of state for $\Lambda$CDM is shown as the horizontal blue line with constant $w=-1$. We compare the curves to the PADE-$w$ parametrization \cite{Alho:2024qds} given by
\begin{equation}
    \epsilon(z) = \frac{3}{2}(1+w(z)) = \frac{3\epsilon_0}{3+\eta_0(z^3+3z^2+3z)}\,. 
\end{equation}
 This parametrization was found by Ref.\ \cite{Shlivko:2025fgv} to accurately reproduce the expansion history for a wide range of thawing-quintessence models. We see that it provides a good approximation for our models.  (Ref.\ \cite{Shlivko:2025fgv} considered only the Union3 SN dataset in their analysis.)  We further contrast our results to the widely used $w_0$-$w_a$  
 CPL parametrization \cite{Chevallier:2000qy, PhysRevLett.90.091301} and the Barboza-Alcaniz parametrization \cite{Barboza:2008rh} that was used by the DESI collaboration to demonstrate evidence for dynamical dark energy \cite{DESI:2025fii}. One can see that these parametrizations do not accurately reproduce the behavior of thawing quintessence fields. Since we are assuming conventional models with a positive kinetic term, we do not encounter phantom crossing of the equation of state ($w<-1$) in the present work.

\section{Discussion}
\label{sec:disc}
We have investigated the assertion that the recent DESI-BAO results favor dynamical dark energy models over the widely accepted cosmological constant. We considered simple quintessence models of a scalar field rolling down an approximately linear or quadratic potential and used data from Planck CMB, DES clustering, DESI 2025 BAO, in addition to SN data from Pantheon+, DESY5 and Union3. The evidence for dynamical dark energy depends strongly on the choice of the SN dataset. With the Pantheon+ or Union3 data, the improvement in $\chi^2$ is less than 4, which is not statistically significant for a model with two extra parameters compared to $\Lambda$CDM. With the DESY5 dataset we find substantial evidence for quintessence, at the cost of moderately exacerbating the Hubble tension. Refs.\ \cite{Ye:2025ark, Cortes:2025joz, Wang:2025bkk} have also noted a significant tension between the different datasets.  This casts some doubt as to whether the claims of evidence for dynamical dark energy can be considered reliable with current data.

This work was partly motivated by the claims of Refs.\ \cite{VanRaamsdonk:2023ion, VanRaamsdonk:2025wvj} that the universe may have already stopped accelerating. That work used similar linear and quadratic approximations to the scalar potential to demonstrate a strong preference for evolving dark energy that eventually becomes negative in the future. However, their analysis omitted CMB data from Planck and clustering data from DES. We find that when only considering BAO and SN data, all three supernova datasets show a stronger preference for quintessence ($\Delta\chi^2 \lesssim -5, -12, -8$ for Pantheon+, DESY5 and Union3 respectively), but upon including Planck data this preference is diminished as we just showed.  Furthermore, these studies imposed final velocities for the quintessence field and integrated the equations of motion backward in time, which typically does not lead to solutions in which $\phi$ is frozen at early times.  Instead, very large values of $\dot\phi$ in the past may be required to get such solutions, which do not correspond to our assumption that $\phi$ was frozen by Hubble damping at early times.

We found that many parametrizations of $w(z)$ that
 are commonly used to test dynamical dark energy do not accurately reproduce the 
predictions corresponding to quintessence models.  Ref.\  \cite{Akthar:2024tua} proposed a more general parametrization of $w(z)$ that can mimic all classes of quintessence dynamics. Their results also indicate that observational data prefer the $\Lambda$CDM model over other models in the non-phantom regime. More recently, Refs.\ \cite{Payeur:2024dnq, Bayat:2025xfr, Dinda:2024ktd, Wang:2025vfb} have concluded that the evidence for dynamical dark energy is marginal.  While the suggestion remains intriguing, we would like to see greater consistency across different data sets before concluding that $w$ is evolving.

\medskip

\textbf{Acknowledgments.} We thank Cliff Burgess, Keshav Dasgupta,  Guillaume Payeur, Fernando Quevedo, and Robert Brandenberger for their helpful comments. We also thank Juan Gallego for assistance with the computing cluster. JC was supported by NSERC (Natural Sciences and Engineering Research Council,
Canada).

\bibliographystyle{utphys}
\bibliography{references}% Produces the bibliography via BibTeX.

\end{document}